\documentclass[pre,showpacs,preprintnumbers,amsmath,amssymb,aps,twocolumn,superscriptaddress]{revtex4}

\usepackage{amsfonts}
\usepackage{bm}
\usepackage{graphicx}
\usepackage{color}
\usepackage{multirow}
\usepackage{epsfig}

\newcommand{\figref}[1]{Fig.\ \ref{#1}}

\newcommand{\neweqnline}{\nonumber\\}
\newcommand{\punc}[1]{\,#1}
\newcommand{\vect}[1]{\mathbf{#1}}

\newcommand{\rs}{r_{\mathrm{s}}}
\newcommand{\Eloc}{E_\mathrm{L}}                 

\newcommand{\ncorr}{n_\mathrm{corr}}             
\newcommand{\PsiT}{\Psi_\mathrm{T}}              
\newcommand{\Rvec}{\vect{R}}                     

\newcommand{\tcorr}{t_\mathrm{corr}}             

\newcommand{\taumax}{\tau_\mathrm{max}}          

\newcommand{\Titer}{T_\mathrm{iter}}             
\newcommand{\nind}{\nu}               

\begin{document}
\title{Strategies for improving the efficiency of quantum Monte Carlo
  calculations} \author{R.\ M.\ Lee}
\affiliation{Theory of Condensed Matter Group, Department of Physics,
  Cavendish Laboratory, 19 J.J.\ Thomson Avenue, Cambridge, CB3 0HE,
  UK} \author{G.\ J.\ Conduit} \affiliation{Department of Condensed
  Matter Physics, Weizmann Institute of Science, Rehovot 76100,
  Israel} \affiliation{Physics Department, Ben Gurion University, Beer Sheva
  84105, Israel} \author{N.\ Nemec} 
 \affiliation{Theory of Condensed Matter Group, Department of Physics,
  Cavendish Laboratory, 19 J.J.\ Thomson Avenue, Cambridge, CB3 0HE,
  UK}\affiliation{Department of Earth Sciences, University College London,
  Gower Street, London WC1E 6BT, UK}
\author{P.\ L\'opez R\'ios} 
\affiliation{Theory of Condensed Matter Group, Department of Physics,
  Cavendish Laboratory, 19 J.J.\ Thomson Avenue, Cambridge, CB3 0HE,
  UK}
\author{N.\ D.\ Drummond}
\affiliation{Theory of Condensed Matter Group, Department of Physics,
  Cavendish Laboratory, 19 J.J.\ Thomson Avenue, Cambridge, CB3 0HE,
  UK} \affiliation{Department of Physics, Lancaster University,
Lancaster LA1 4YB, UK}\date{\today}

\begin{abstract}
We describe a number of strategies for minimizing and calculating 
accurately
the statistical uncertainty in quantum Monte Carlo calculations. We
investigate the impact of the sampling algorithm on the efficiency 
of the
variational Monte Carlo method. Our finding that a relative 
time-step ratio
of $1:4$ is optimal in DMC is in agreement with the result of 
[J.~Vrbik and
S.~M.~Rothstein, Intern. J. Quantum Chem. 29, 461-468 (1986)]. 
Finally, we
discuss the removal of serial correlation from data sets by 
reblocking,
setting out criteria for the choice of block length and 
quantifying the
effects of the uncertainty in the estimated correlation length.  
\end{abstract}

\maketitle


\section{Introduction\label{sec:intro}}
Quantum Monte Carlo (QMC) methods are a class of stochastic \textit{ab
  initio} techniques for solving the many-body Schr\"{o}dinger
equation~\cite{foulkes-qmcsos2001,needs-cvadqmcc2010}. They are
capable of achieving accuracy comparable to that of post-Hartree-Fock
quantum-chemistry techniques but with a much lower computational
cost. The diffusion Monte Carlo (DMC) method in particular has no
close competitors for calculations of the energy of bulk periodic
systems. 

The utility of QMC stems from the fact that the cost of achieving a
given error bar scales as $\sim N^3$ for typical
systems~\footnote{Ultimately, for large $N$ the scaling of DMC with
  system size becomes exponential as discussed in N.\ Nemec, Phys.\
  Rev.\ B \textbf{81}, 035119 (2010).}, where $N$ is the number of
quantum particles. The method is most useful when studying systems for
which quantum-chemistry calculations are infeasible and density
functional theory does not give a sufficiently accurate description of
electronic correlation. The algorithms are intrinsically parallel,
allowing QMC to take full advantage of developments in computer
technology. The variational Monte Carlo (VMC) algorithm, for example,
is almost perfectly parallelizable.  Furthermore, existing QMC
implementations are easily extended to different systems. One may
apply the same basic algorithms, changing only the form of the trial
wave function and the Hamiltonian, to systems comprising any
combination of particles and interparticle interactions.  Because the
trial wave function can be an explicit function of interparticle
distances, the Kato cusp conditions and other correlation effects can
be described compactly, without the need for large expansions of many
determinants and other unwieldy functional
forms~\cite{kato1957eigenfunctions}. For a comprehensive overview of
VMC and DMC, the reader is directed to Refs.\
\onlinecite{needs-cvadqmcc2010,foulkes-qmcsos2001,casino2009,umrigar-admcawvste1993}.


The practical challenges facing QMC are largely concerned with
improving the efficiency of the algorithms and the design of new trial
wave functions. The computational expense of a large calculation
necessitates careful selection of the operational
parameters. Typically one has a certain amount of computer time
available within which one wishes to achieve the smallest possible
statistical error in the final result. In addition, the extraction of
an accurate statistical error bar from serially correlated data is
itself nontrivial. In this paper, we outline how to choose the
optimal parameters and algorithms at the different stages of a QMC
calculation and describe how to process the resulting data.

This paper is structured as follows. Section \ref{sec:VMC_eff} gives
an analysis of the many related factors contributing to the efficiency
of VMC calculations. Section \ref{sec:extrapolation} describes how to
improve the efficiency of DMC time step extrapolation. In Sec.\
\ref{sec:reblocking} we discuss the calculation of accurate error bars
using the reblocking method and describe a robust scheme for choosing
block lengths. We demonstrate in Sec.\ \ref{sec:outliers} that
uncertainty in the estimated correlation length results in an error in
the statistical error bar that can significantly enhance the
probability of observing outliers. Finally, we draw our conclusions in
Sec.\ \ref{sec:conclusions}.  We use Hartree atomic units
($\hbar=|e|=m_{\rm e}=4\pi\epsilon_0=1$) throughout this article.

\section{Efficiency of VMC calculations \label{sec:VMC_eff}}
\subsection{Method}

In this section, we discuss practical schemes for achieving maximal
efficiency within the VMC method. We focus on three aspects of a VMC
calculation. The first is the sampling algorithm, which is how moves
are proposed. The second is the use of decorrelation loops, which
consist of additional moves for which we avoid evaluating the local
energy. We will demonstrate that decorrelation loops can offer a
twofold increase in efficiency. To our knowledge, there are no
quantitative investigations of decorrelation loops in the
literature. The third factor we consider is the choice of time step,
which governs the width of the transition probability density function
(PDF)\@. Our findings are summarized by the set of recommendations in
Sec.\ \ref{sec:rec_vmc}.

Variational Monte Carlo is the simplest and least computationally
expensive QMC method. In the VMC method, the expectation value of the
Hamiltonian $\hat{H}$ with respect to a trial wave function
$\Psi_{\mathrm{T}}$ is calculated using a stochastic integration
technique, giving a variational estimate for the ground state energy,
\begin{equation}
  \frac{\langle \PsiT | \hat{H} | \PsiT \rangle}{\langle \PsiT
|\PsiT \rangle}
  = \frac{\int{\rm d}\Rvec |\PsiT(\Rvec)|^2 \Eloc(\Rvec)}{
    \int{\rm d}\Rvec |\PsiT(\Rvec)|^2}
  \approx \frac{1}{n}\sum_{i=1}^{n}E_i \;,
\label{eq:vmc_integral}
\end{equation}
where $\Eloc(\Rvec)=\PsiT^{-1}(\Rvec)\hat{H}\PsiT(\Rvec)$ is the local
energy and $\vect{R}$ is a vector describing all the particle
positions. The set $\lbrace E_i \rbrace_{i=1,\ldots,n}$ contains $n$
energies and is produced by evaluating $E_i=\Eloc(\Rvec_i)$ at $n$
points $\{\Rvec_i\}_{i=1,\ldots,n}$ in configuration space distributed
according to $|\PsiT(\Rvec)|^2$.

Due to the finite number of samples $n$, the VMC estimate of the
energy of Eq.\ (\ref{eq:vmc_integral}) has a statistical error
{$\Delta_0 = \sigma_0(n/\ncorr)^{-1/2}$}, where $\sigma_0$ is the
standard deviation of the local-energy distribution and $\ncorr$ is
the correlation length~\cite{wolff-mcewle2004} of the sequence of
local energies.

The quantity $\sigma_0$ only depends on the system and the trial wave
function, whereas $\ncorr$ also depends on the sampling algorithm.
Thus, for a given system, trial wave function, and sampling algorithm,
the statistical error diminishes with the number of configurations
sampled as $n^{-1/2}$.  Suppose one VMC step takes a time $\Titer$. A
VMC calculation is more efficient the less time it requires to reach a
given statistical error $\Delta_0$, so if a VMC run takes a CPU time
of $T = n \Titer$ to sample $n$ configurations, an appropriate measure
of its efficiency is
\begin{equation} {\cal E} = \left( \Delta_0^2 n \Titer
\right)^{-1} = \left( \sigma_0^2 \ncorr \Titer \right)^{-1}\;,
\label{eq:vmc_efficiency}  \end{equation}
which is independent of $n$. The efficiency of a VMC calculation can
be improved by reducing the product $\ncorr \Titer$.

\subsection{VMC sampling}
The electronic configurations $\{\Rvec_i\}_{i=1,\ldots,n}$ are
generated using the Metropolis
algorithm~\cite{metropolis-eoscbfcm1953}, where a move from $\Rvec_i$
to $\Rvec_i^\prime$ is proposed with probability $T(\Rvec_i^\prime
\leftarrow \Rvec_i)$, and is accepted (\textit{i.e.},
$\Rvec_{i+1}=\Rvec_i^\prime$) with probability
\begin{equation}
\label{eq:vmc_acceptance}
A(\Rvec_i^\prime \leftarrow \Rvec_i) =
\min\left(
  1,
  \frac
    {T(\Rvec_i \leftarrow \Rvec_i^\prime)}
    {T(\Rvec_i^\prime \leftarrow \Rvec_i)}
  \frac
    {\left|\PsiT(\Rvec_i^\prime)\right|^2}
    {\left|\PsiT(\Rvec_i)\right|^2}
\right)
\;,
\end{equation}
or otherwise rejected (\textit{i.e.}, $\Rvec_{i+1}=\Rvec_i$). 
In fact, if the wave function can be factorized, one can greatly
improve efficiency using multi-level
sampling~\cite{dewing_twolevel}. All of our calculations use two-level
sampling, in which we accept or reject the move first based on the
Slater determinant part of $\PsiT(\Rvec)$ and then (if the Slater part
of the move was accepted) based on the Jastrow
factor~\cite{foulkes-qmcsos2001}.


A simple, commonly used choice for $T(\Rvec_i^\prime \leftarrow
\Rvec_i)$ is the product of Gaussian distributions of variance $\tau$
(standard deviation $\sqrt{\tau}$) for each of the Cartesian
components of the displacement of each electron.  By analogy with DMC,
$\tau$ is often referred to as the VMC ``time step,'' although there
is no notion of \textit{time} in the VMC formalism. We shall restrict
our analysis to the case of Gaussian transition
probabilities. Alternatives to this choice have been
proposed~\cite{umrigar_transprob,stedman_transprob}, but these studies
focus on the statistical improvement for a given number of iterations,
and do not analyze the total efficiency.  The simplicity of the
Gaussian distribution represents an efficiency advantage that is hard
to offset with more exotic distributions.  Nonetheless, the
conclusions presented here should mostly be applicable to other
transition probabilities.

\subsubsection{Configuration-by-configuration and electron-by-electron
sampling} In the sampling algorithm we have just described, to go from
$\Rvec_i$ to $\Rvec_{i+1}$ we propose an entire configuration move, and we
accept it or reject it with a single decision.  This is what we call
configuration-by-configuration sampling (CBCS)\@.

However, it is possible to generate $\Rvec_{i+1}$ from $\Rvec_i$ by
proposing $N$ successive single-electron moves and accepting or
rejecting each of them individually.  The resulting algorithm is
electron-by-electron sampling (EBES), which allows larger moves to be
accepted, greatly reducing $\ncorr$.  This comes at the cost of an
increase in $\Titer$, because evaluating the $N$ acceptance
probabilities in EBES takes longer than computing the single
acceptance probability in CBCS\@.

\subsubsection{Averaging local energies over proposed moves}

It is possible to replace the average in Eq.\ (\ref{eq:vmc_integral})
with an expression where the local energies at $\Rvec_i^\prime$ and
$\Rvec_i$ are multiplied by the acceptance and rejection
probabilities, respectively, and summed together. For CBCS, the
expression is~\cite{ceperley_ancientvmc}:
\begin{eqnarray}
\frac
  {\langle \PsiT | \hat{H} | \PsiT \rangle}
  {\langle \PsiT | \PsiT \rangle}
&\approx&
\frac{1}{n}\sum_{i=1}^{n}
\left\{
  A(\Rvec_i^\prime\leftarrow\Rvec_i)\Eloc(\Rvec_i^\prime)
  \right. \nonumber\\ &+& \left.
  \left[ 1 - A(\Rvec_i^\prime\leftarrow\Rvec_i) \right] \Eloc(\Rvec_i)
\right\}
\;.
\end{eqnarray}
This expression is also a valid approximation to the VMC energy, with
the advantage that rejected moves contribute to the sum, adding new
data and improving the statistics, especially when the acceptance
ratio is low.  This translates into a reduction in $\ncorr$.  The
evaluation of the additional local energies increases $\Titer$,
however.  We investigate the balance of these factors below for
CBCS\@.

We have avoided averaging the energy over proposed moves in EBES since
even with refinements it has been found to be less efficient than the
unmodified algorithm~\cite{ndd_thesis}.


\subsubsection{Decorrelation loops\label{sec:decorr_loops}}
It is possible to go from $\Rvec_i$ to $\Rvec_{i+1}$ by proposing $p$
configuration moves in turn instead of just one.  In this scheme one
generates a sample of $n$ local energies by performing a calculation
consisting of $pn$ moves and evaluating the local energy at every
$p$th configuration.

The cost of one step of a VMC calculation with a decorrelation loop of
length $p$ is
\begin{equation}
\label{eq:time_p}
\Titer(p) = p T_{\rm move} + T_{\rm energy}
\;,
\end{equation}
where $T_{\rm move}$ is the time it takes to propose and accept or
reject a single configuration move and $T_{\rm energy}$ is the time it
takes to evaluate the local energy~\footnote{The details of the
  implementation may need to be taken into account in this expression.
  An implementation could detect whether all moves have been rejected
  between evaluations of the local energy to avoid unnecessary
  re-evaluations.  In this case, the probability of not having to
  calculate a local energy is the probability of having rejected $p$
  consecutive moves, and $\Titer(p)$ becomes $p T_{\rm move} + \left[
    1 - \left( 1 - a \right)^p \right] T_{\rm energy}$.}.

It is possible to establish the precise form of the correlation length
$\ncorr(p)$ as a function of $p$.  When $n\rightarrow\infty$,
$\ncorr$ is
\begin{equation}
\label{eq:ncorr_infty}
\ncorr = \ncorr(1) = 1 + 2 \sum_{k=1}^\infty {\cal A}_k \;,
\end{equation}
where ${\cal A}_k$ is the autocorrelation of local energies separated by $k$
steps,
\begin{equation}
  {\cal A}_k = \frac{1}{\sigma_0^2}
   \Big \langle
(E_l-\langle E\rangle)(E_{k+l}-\langle E \rangle) \Big \rangle_l
  \;.
\end{equation}
If we assume that the autocorrelation is dominated by a single exponential
term, \textit{i.e.}, ${\cal A}_k=\exp(-\alpha k)$ then Eq.\ (\ref{eq:ncorr_infty})
becomes
\begin{equation}
\ncorr =
1 + 2 \sum_{k=1}^\infty \exp(-\alpha k) =
1 + 2 \frac {\exp(-\alpha)} {1-\exp(-\alpha)} \;.
\end{equation}
Hence $\exp(-\alpha) = (\ncorr-1) / (\ncorr+1)$,
and the correlation length at $p$ is
\begin{eqnarray}
\ncorr(p) & = & 1 + 2 \sum_{k=1}^\infty {\cal A}_{pk} \nonumber\\ & = & 1 + 2
                \frac {\left(\ncorr-1\right)^p} {\left(\ncorr+1\right)^p-
                \left(\ncorr-1\right)^p} \;, \label{eq:ncorr_p_infty}
\end{eqnarray}
which falls off as $p^{-1}$ if $\ncorr$ is large.
From Eqs.\ (\ref{eq:vmc_efficiency}), (\ref{eq:time_p}), and
(\ref{eq:ncorr_p_infty}) we can build the full expression for ${\cal E}$, and
it is possible to find the value of $p$ that maximizes ${\cal E}$ analytically
from estimates of $T_{\rm move}$, $T_{\rm energy}$, and $\ncorr$.

The usefulness of decorrelation loops depends on how costly it is to
evaluate local energies and how much serial correlation is present.
Were it the case that local energies took no time to evaluate (\textit{i.e.},
$T_{\rm energy}=0$), the inclusion of decorrelation loops would not
increase the efficiency ${\cal E}$, and if no serial correlation were
present then $\ncorr(p) = 1$, and increasing $p$ would simply increase
the cost of each step.

\subsection{Automatic optimization of $\tau$}

Although the VMC algorithm is valid for any positive time step, the efficiency
of the method depends strongly on $\tau$.  An appropriate time step for EBES
VMC can be very roughly estimated as being such that the root-mean-square
(RMS) distance moved by each electron at each time step is equal to the most
important physical length scale in the problem.  Assuming the acceptance
probability of electron moves is approximately 50\%, the RMS distance diffused
is $\sqrt{3 \tau /2}$ in three dimensions.  In an electron gas the only
physical length scale is the radius $\rs$ of the sphere that contains one
electron on average, so the required time step is $\tau \approx 2\rs^2/3$.  In
an atom the length scale is somewhere between the Bohr radius $1/Z$, where $Z$
is the atomic number, and 1 a.u.  However, it is clear that these crude
choices are far from optimal.

There are two commonly-used approximate methods for choosing $\tau$;
aiming to achieve an acceptance ratio of 50\% (the 50\% rule), and
maximizing the diffusion constant. Both can be implemented so that
this optimization occurs automatically and inexpensively at the
beginning of a VMC run.

In the ``50\% rule'' it is assumed that the ratio $a$ of accepted
moves to proposed moves is representative of the sampling efficiency,
and that a value of 50\% is near-optimal. In general, the two limits
of 0\% and 100\% acceptance correspond to a failure to properly
explore phase space, but there is no particular reason why $a=50\%$
should correspond to optimal sampling.

The diffusion constant $D$ can be computed as the average of the
squared displacement between consecutive configurations $\Rvec_i$ and
$\Rvec_{i+1}$~\footnote{When we use decorrelation loops, we define $D$
  as the average of the squared displacement between consecutive
  configurations within the decorrelation loop, not between those for
  which the energy is evaluated.}. One might reasonably assume that
choosing $\tau$ to maximize $D$ is an efficient strategy, although
maximization of $D$ does not necessarily correspond to optimal
sampling. For example, in a CBCS study of the homogeneous electron
gas, rigidly translating all of the electrons together results in a
very large diffusion constant but clearly corresponds to poor
exploration of phase space.

\subsection{Empirical data and analysis \label{sec:empirical_vmc}}

We shall consider four basic choices to be made when performing a VMC
calculation with a Gaussian transition-probability density: whether to
use CBCS or EBES, whether to average local energies over proposed
moves, the value of the ``time step'' $\tau$, and the length of the
decorrelation loop $p$.

In order to study the effect of these choices, we have performed VMC
calculations for a set of six representative systems: a
pseudopotential N atom, an all-electron O atom, a pseudopotential NiO
molecule, an all-electron N${}_2$H${}_4$ molecule, a three-dimensional
homogeneous electron gas (HEG) composed of 38 electrons at a density
parameter of $\rs=1$ a.u., and a 16-atom supercell of a
pseudopotential C diamond crystal \footnote{Our calculations were
  performed on a cluster of eight 24GB, dual-socket, quad-core,
  2.66GHz Intel Core i7 processors.  However, we have only quoted
  ratios of efficiencies in this paper, which should be largely
  architecture-independent.}. For each system, we tested two trial
wave functions: one of Slater-Jastrow
form~\cite{foulkes-qmcsos2001,drummond_jastrow} and another of
Slater-Jastrow-backflow form~\cite{kwon_backflow,plr_backflow}.

\begin{table}[!ht]
\begin{center}
\begin{tabular}{l @{~} r@{.}l @{~} r @{~} r @{~} r@{.}l @{~}
                r@{.}l @{~} r@{.}l @{~} r@{.}l}
\hline \hline
System                                                          &
\multicolumn{2}{c}{$\tau_{\rm opt}$}                            &
\multicolumn{1}{c}{$p_{\rm opt}$}                               &
\multicolumn{1}{c}{$a_{\rm opt}$}                               &
\multicolumn{2}{c}{${\cal E}_{50\%}/{\cal E}_{\rm opt}$}        &
\multicolumn{2}{c}{${\cal E}_{D_{\rm max}}/{\cal E}_{\rm opt}$} &
\multicolumn{2}{c}{${\cal E}_{p=1}/{\cal E}_{\rm opt}$}         \\
\hline
N (pp)         & ~~0&20 & 3~ & ~55\% & ~~~~1&00 & ~~~~0&52 & ~~~~0&65 \\
O              &   0&05 & 3~ &  58\% &     0&94 &     0&40 &     0&68 \\
NiO (pp)       &   0&20 & 5~ &  44\% &     0&94 &     0&39 &     0&41 \\
N${}_2$H${}_4$ &   0&05 & 3~ &  64\% &     0&62 &     0&09 &     0&71 \\
HEG            &   1&00 & 3~ &  37\% &     0&93 &     0&96 &     0&73 \\
Diamond        &   1&00 & 3~ &  32\% &     0&94 &     0&66 &     0&60 \\
\hline \hline
\end{tabular}
\caption{Optimal parameters and comparison of different efficiencies
for EBES using Slater-Jastrow wave functions. Pseudopotentials (pp)
were used in some of the calculations.
\label{table:ebes_data}}
\end{center}
\end{table}
\begin{table}[!ht]
\begin{center}
\begin{tabular}{l @{~} r@{.}l @{~} r @{~} r @{~} r@{.}l @{~}
                r@{.}l @{~} r@{.}l @{~} r@{.}l}
\hline \hline
System                                                          &
\multicolumn{2}{c}{$\tau_{\rm opt}$}                            &
\multicolumn{1}{c}{$p_{\rm opt}$}                               &
\multicolumn{1}{c}{$a_{\rm opt}$}                               &
\multicolumn{2}{c}{${\cal E}_{50\%}/{\cal E}_{\rm opt}$}        &
\multicolumn{2}{c}{${\cal E}_{D_{\rm max}}/{\cal E}_{\rm opt}$} &
\multicolumn{2}{c}{${\cal E}_{p=1}/{\cal E}_{\rm opt}$}         \\
\hline
N (pp)         & ~~0&10 &  8~ & ~32\% & ~~~~0&87 & ~~~~0&90 & ~~~~0&33 \\
O              &   0&01 &  8~ &  27\% &     0&82 &     0&82 &     0&64 \\
NiO (pp)       &   0&02 & 36~ &  17\% &     0&70 &     1&00 &     0&13 \\
N${}_2$H${}_4$ &   0&01 & 13~ &  16\% &     0&59 &     1&00 &     0&42 \\
HEG            &   0&05 & 36~ &   9\% &     0&54 &     0&88 &     0&47 \\
Diamond        &   0&02 & 36~ &  11\% &     0&25 &     0&79 &     0&13 \\
\hline \hline
\end{tabular}
\caption{Optimal parameters and comparison of different efficiencies
for CBCS using Slater-Jastrow wave functions.
\label{table:cbcs_data}}
\end{center}
\end{table}
For each system and wave function, we have performed calculations using EBES
and CBCS, and for CBCS we have run calculations with and without averaging
over proposed moves.  Finally, for each system, wave function, and sampling
method, we have performed 160 VMC calculations covering 16 different values of
$\tau$ and 10 different values of $p$. In each case we have identified the
maximum efficiency ${\cal E}_{\rm opt}={\cal E}(\tau_{\rm opt},p_{\rm opt})$.
To assess the performance of the ``50\% rule,'' we have located the value of
the time step $\tau_{50\%}$ whose acceptance ratio is closest to $50\%$ and
compared the efficiency ${\cal E}_{\rm 50\%}= {\cal E}(\tau_{\rm 50\%},p_{\rm
opt})$ with ${\cal E}_{\rm opt}$.  To assess the performance of maximizing the
diffusion constant, we have located the value of the time step $\tau_{D_{\rm
max}}$ with the maximum $D$ and compared the efficiency ${\cal E}_{D_{\rm
max}}={\cal E}(\tau_{D_{\rm max}}, p_{\rm opt})$ with ${\cal E}_{\rm opt}$.
To assess the importance of decorrelation loops, we have compared the
efficiency ${\cal E}_{p=1}={\cal E}(\tau_{\rm opt},1)$ with ${\cal E}_{\rm
opt}$.  The results of these comparisons are given in Table
\ref{table:ebes_data} for EBES and Table \ref{table:cbcs_data} for CBCS, in
both cases for the Slater-Jastrow wave function only; the data for the
Slater-Jastrow-backflow wave function are nearly identical and are not shown.

For the periodic systems the acceptance ratio in EBES does not reach
zero as $\tau$ is increased, and as a consequence the efficiency
presents a plateau in that region, where we find that ${\cal E}$ is
close to ${\cal E}_{\rm opt}$.
In EBES we also find that the ``50\% rule'' consistently gives
efficiencies within 10\% of the maximum, with the exception of the
N${}_2$H${}_4$ molecule, where the optimal acceptance ratio is larger.
Maximization of the diffusion constant in EBES consistently gives time
steps that are too large and yields efficiencies below about 50\% of
the maximum possible for finite systems, and between 65\% and 95\% of
the maximum for periodic systems.  In CBCS, maximizing the diffusion
constant achieves reasonable efficiencies, often within 10\% of the
maximum value, while the ``50\% rule'' gives increasingly poor results
as the system size increases. Decorrelation loops improve the
efficiency in EBES by between 50\% and 150\%.  In CBCS these become
more important and enhance ${\cal E}$ by up to a factor of seven.

\begin{table}[!ht]
\begin{center}
\begin{tabular}{l@{~~~}c@{~~~}r@{.}lr@{.}l@{~~~}c@{~~~}r@{.}lr@{.}l}
  \hline \hline
  & & \multicolumn{4}{c}{${\cal E}_{\rm EBES}/{\cal E}_{\rm CBCS}$} & &
  \multicolumn{4}{c}{${\cal E}_{\rm CBCS}/{\cal E}_{\rm CBCS2}$} \\
  \raisebox{1.5ex}[0ex]{System}    &
  \raisebox{1.5ex}[0ex]{$N$}       &
  \multicolumn{2}{c}{SJ}  &
  \multicolumn{2}{c}{SJB} & &
  \multicolumn{2}{c}{SJ}  &
  \multicolumn{2}{c}{SJB} \\
  \hline
  N (pp)         & ~5 &          ~1&05 &        ~~~0&90   & &   ~1&22 &        ~~~1&24 \\
  O              & ~8 &           1&47 &           1&07   & &    1&10 &           1&17 \\
  NiO (pp)       & 16 &           1&65 &           1&22   & &    1&38 &           1&52 \\
  N${}_2$H${}_4$ & 18 &           1&93 &           0&83   & &    1&11 &           1&53 \\
  HEG            & 38 &           3&11 &           1&95   & &    1&27 &           1&25 \\
  Diamond        & 64 &           4&70 &           2&36   & &    1&14 &           1&24 \\
  \hline \hline
\end{tabular}
\caption{Comparison of the efficiency of EBES and CBCS for
  Slater-Jastrow (SJ) and Slater-Jastrow-backflow (SJB) wave
  functions, and also for averaging local energies over
  proposed moves (CBCS2) and computing a single energies (CBCS)\@.
  \label{table:ebes_vs_cbcs}}
\end{center}
\end{table}
In Table \ref{table:ebes_vs_cbcs} we compare the maximum efficiency
encountered in EBES ${\cal E}_{\rm EBES}$ with that in CBCS ${\cal
  E}_{\rm CBCS}$ for Slater-Jastrow (SJ) and Slater-Jastrow-backflow
(SJB) wave functions. The fifth and sixth columns of Table
\ref{table:ebes_vs_cbcs} show the comparison for CBCS when a single
energy is evaluated per configuration move (${\cal E}_{\rm CBCS}$),
and where averages of local energies over proposed moves are carried
out (${\cal E}_{\rm CBCS2}$).

EBES is more efficient in all cases, with the exception of the
backflow calculations on the pseudopotential N atom and the
all-electron N${}_2$H${}_4$ molecule.  The improvement in efficiency
that EBES offers over CBCS increases with system size. Averaging
energies over proposed moves is found to be less efficient in every
case.
\subsection{Recommendations\label{sec:rec_vmc}}

Our key finding is that decorrelation loops increase the efficiency of
EBES by roughly a factor of two and that of CBCS by much more. One can
use the expressions in Sec.\ \ref{sec:decorr_loops} to determine the
optimal loop length $p$, although in practice a decorrelation period
of $p=3$ delivers near-optimal efficiency in the EBES algorithm for a
wide range of systems.

Based on the data presented in Sec.\ \ref{sec:empirical_vmc}, we
suggest that EBES should nearly always be used in VMC, the only
possible exception being for small systems with fewer than about 20
electrons when backflow is used.  (Even in this case, CBCS is not much
more efficient than EBES\@.)  When using EBES, one should use the
``50\% rule'' to optimize the time step $\tau$.  
If CBCS is used, one should maximize the diffusion constant to
optimize the time step $\tau$.
Finally, we find that accumulation methods which average local
energies over proposed moves are less efficient for every system
tested.



\section{Optimizing DMC time-step
  extrapolation \label{sec:extrapolation}}

DMC is a Green's function projector method for solving the
Schr\"{o}dinger equation in imaginary time. In DMC, the ground state
distribution is represented by the density of walkers (points in
configuration space) rather than by an analytic function. Propagation
of a population of walkers in imaginary time projects out the
ground-state component of the initial DMC wave
function~\cite{ceperley-gsotegbasm1980,foulkes-qmcsos2001}.

The DMC algorithm is only accurate in the limit of small time step
$\tau$. However, the computational effort required to achieve a given
error bar goes as $1/\tau$, ruling out the use of infinitesimal time
steps in practice.  Hence, where high accuracy is required, two or
more finite time steps $\{\tau_i\}$ are generally used and the
ground-state energy is obtained by extrapolating to
$\tau=0$~\cite{foulkes-qmcsos2001,needs-cvadqmcc2010}.  Here we
explain how the statistical error in a zero-time-step extrapolate may
be minimized by a judicious choice of time steps $\{\tau_i\}$, and the
prudent deployment of a limited total computing time between those
time steps. Note: since our paper was published, it has been drawn to 
our attention that the principal result obtained in this section was 
previously derived in Ref.\ \onlinecite{Rothstein}.

For sufficiently small $\tau$, the DMC energy scales linearly with the
time step as $E(\tau)=E_0+\kappa \tau$.  Suppose we calculate
$E(\tau)$ at $R$ different time steps $\lbrace\tau_i\rbrace$ in the
linear-bias regime, where each $E(\tau_i)$ has an associated
statistical uncertainty $\Delta_i$. The error bars fall off with the
time step $\tau_i$ and the CPU time devoted to the calculation $T_i$
as $\Delta_i=C/\sqrt{\tau_i T_i}$, where $C$ is a constant.  To
determine the ground-state energy at zero time step $E_0$, we minimize
the $\chi^2$ error of the linear fit,
\begin{eqnarray}
  \chi^2 & = & \sum_{i=1}^R \frac{[E(\tau_i)-E_0- \kappa
    \tau_i]^2}{\Delta_i^2}\neweqnline &=&\frac{1}{C^2}\sum_{i=1}^{R}
    T_i \tau_i [E(\tau_i)-E_0- \kappa \tau_i]^2
\end{eqnarray}
with respect to $\kappa$ and $E_0$.  Setting
${\partial\chi^{2}/\partial\kappa=\partial\chi^{2}/\partial E_0=0}$,
we obtain
\begin{equation}
  E_0=\frac{2 \sum_{i=1}^{R}\sum_{j=1}^{R} E(\tau_i)T_i T_j \tau_i \tau_j^2
(\tau_j-\tau_i)}{\sum_{i=1}^R \sum_{j=1}^R T_i T_j \tau_i \tau_j
(\tau_j-\tau_i)^2}\;.
\end{equation}
Assuming the data are Gaussian-distributed, the square of the standard error
in the extrapolate $E_0$ is
\begin{eqnarray}
  \Delta_0^2 & \approx & \sum_{k=1}^R \Delta_k^2 \left[
      \frac{\partial E_0}{\partial E(\tau_k)} \right]^2 \neweqnline
      & = & 4 C^2 \sum_{k=1}^R T_k \tau_k \!\! \left[
      \frac{\sum_{j=1}^R T_j \tau_j^2 (\tau_j - \tau_k)}{\sum_{i=1}^R
      \sum_{j=1}^R T_i T_j \tau_i \tau_j (\tau_j-\tau_i)^2}\right]^2 \punc{.}
\label{eq:delta0}
\nonumber \\
\end{eqnarray}
As expected the standard error falls off as the time steps
$\{\tau_i\}$ are increased and as more time $\{T_i\}$ is dedicated to
the calculations. However, $\tau$ should not be increased beyond
$\tau_{\mathrm{max}}$, the limit of the region in which the bias is
linear. The effort allocated to the calculations cannot be increased
indefinitely because one is constrained by the total time
$T=\sum_{i=1}^R T_i$ for all of the simulations. We now minimize
$\Delta_0^2$ subject to the constraint that $T$ is fixed.

\begin{figure}
 \centerline{\resizebox{0.85\linewidth}{!}{\includegraphics{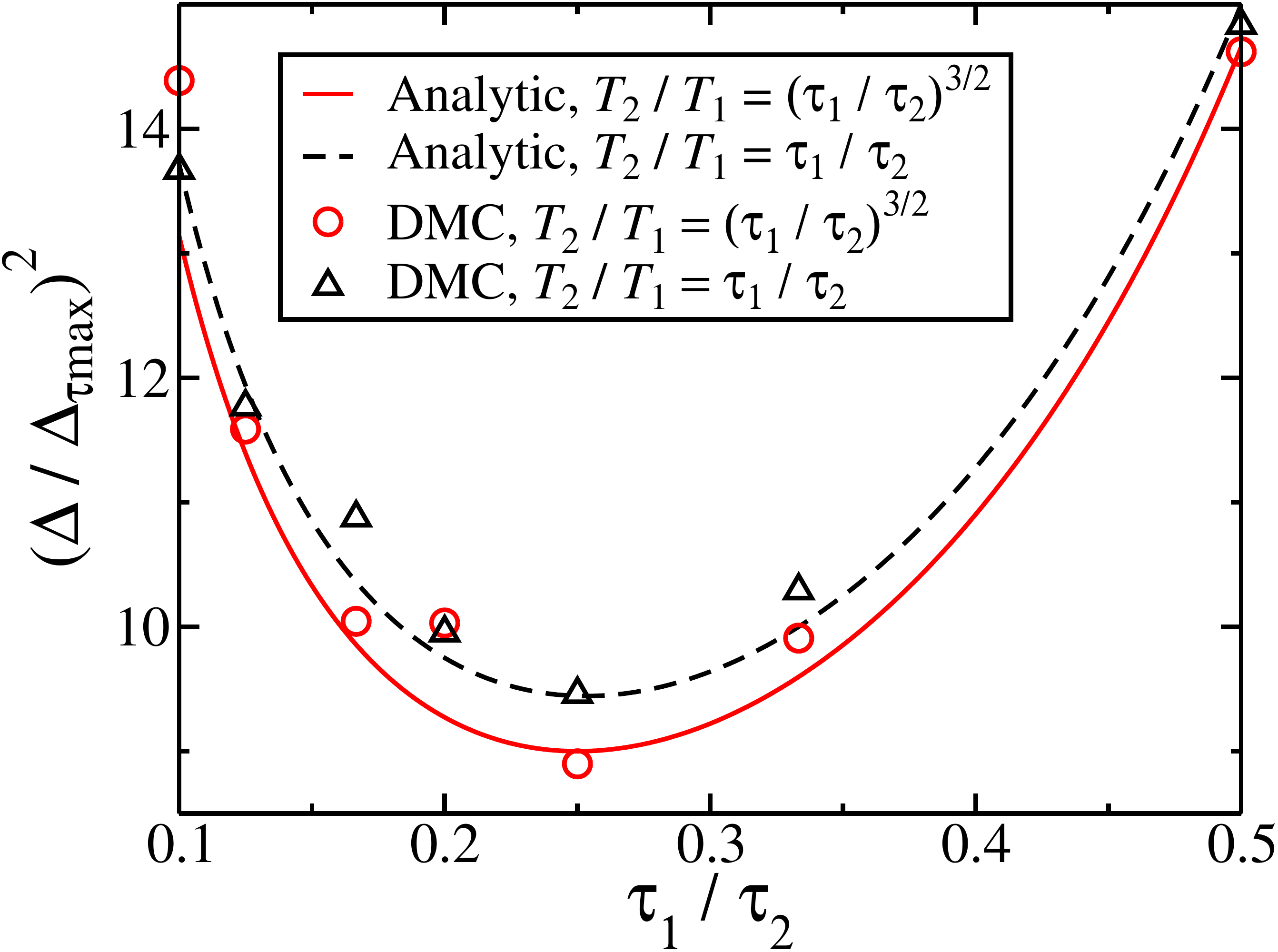}}}
 \caption{(Color online) The uncertainty in the extrapolated DMC
   energy against relative step size, $\tau_1/\tau_2$. The solid line and
   circles show the uncertainty in the extrapolated results obtained
   with the optimal relative run times
   {$[T_1/T_2=(\tau_2/\tau_1)^{3/2}]$}, and the dashed line and the
   triangles with the effort distributed such that the energies have
   equally sized error bars {$(T_1/T_2=\tau_2/\tau_1)$}. The symbols are
   DMC data from the one-dimensional HEG\@. The error bars are
   normalized by $\Delta_{\tau \rm max}$, the error bar of a DMC run
   at the upper time step $\tau_2$ if all of the computational
   resources $(T_1+T_2)$ were dedicated to it.}
 \label{fig:TimestepExtrapolation}
\end{figure}

Let us first suppose that we are to perform just $R=2$ simulations.
We start by fixing the time steps $\tau_1$ and $\tau_2$, and
minimizing $\Delta_0^2$ with respect to the run lengths in the
presence of a Lagrange multiplier to constrain the total run time
$T$. This yields the optimal simulation durations
$T_1=T\tau_2^{3/2}/(\tau_1^{3/2}+\tau_2^{3/2})$ and
$T_2=T\tau_1^{3/2}/(\tau_1^{3/2}+\tau_2^{3/2})$. This deployment
attempts to reduce the error bar on the calculation with the smallest
time step beyond the distribution of effort $T_1/T =
\tau_2/(\tau_1+\tau_2)$ that would ensure error bars of equal
size. Without loss of generality, we now assume that $\tau_2>\tau_1$,
with $\tau_2=\tau_{\mathrm{max}}$ pinned near the boundary of the
linear regime, and we search for the optimal time step $\tau_1$. Using
the optimal durations $T_1$ and $T_2$, minimization of $\Delta_0^{2}$
reveals that the optimal choice of time step is
$\tau_{1}=\tau_{2}/4$. The corresponding optimal physical run times
are therefore $T_{1}=8T/9$ and $T_{2}=T/9$. The full dependence of the
final error upon the relative time step $\tau_{1}/\tau_{2}$ is shown
in \figref{fig:TimestepExtrapolation}.

Now suppose that more than two time steps are used to perform the
extrapolation.  We find that $\Delta_{0}^{2}$ is minimized when all
the computational effort is dedicated to the two points that are
nearest to having a relative time step of $4$ and have the largest
maximum value of $\tau$. Computational effort should therefore be
focused solely on that optimal pair as long as the linear regime is
well-defined. There is thus no advantage to using more than $R=2$ data
points.

%
%

Our scheme is the optimal extrapolation procedure when the extent of
the linear regime is known. The strategy is thus highly applicable to
studies of many similar systems where the linear regime can be assumed
to be the same for multiple runs. For systems where the behavior of
the time step bias has not been established, one has no alternative
but to perform multiple runs over a wide domain of time steps and
determine where the spectrum first increases superlinearly. In such
cases, one can use the RMS distance diffused by an electron over a
single step as an initial order-of-magnitude estimate for where the
linear regime begins. For all-electron atomic systems, for example,
one would expect the linear regime to occur for time steps less than
of the order $\tau=1/(3Z^2)$, where $Z$ is the largest atomic number
occurring in the system. This choice of time step ensures that the RMS
distance diffused is equal to one Bohr radius of the largest atom
under study. For a homogeneous electron gas, where the only
physically-significant length scale is defined by the density, the
equivalent time step would be $\tau=(\rs^2)/d$, where $\rs$ is the
radius of the sphere (circle in 2D) that contains one electron on
average, and $d$ is the dimensionality. Time step bias is reduced when
the modifications of Ref.\ \onlinecite{umrigar-admcawvste1993} are
made to the DMC Green's function, and also when higher-quality wave
functions are used.

If one has accumulated a significant set of results for $\tau<\taumax$
in determining the extent of the linear regime, the prescription for
minimizing the error in the extrapolate has the potential to differ
from the two-run procedure. If one has a large amount of computing
time remaining after determining $\taumax$, the two-run approach is
unchanged. In the event that little computing time remains after
determining $\taumax$, one should devote the remaining time to the run
whose contribution falls the quickest with computer time,
\textit{i.e.,} the run $i$ with the most negative value of $\partial
\Delta_0 / \partial T_{i}$, which may be found from Eq.\
(\ref{eq:delta0}).

Avoiding higher order fitting functions and using only data from
within the linear regime for the extrapolation is the most robust
strategy. Though the formalism here can be extended to study
higher-order fitting functions, finding the appropriate regimes for
higher-order terms would require a larger amount of computational
effort and there is a danger of numerical stability and branching
problems affecting calculations for very large $\tau$. Linear
extrapolation is always an option since the leading-order term in the
bias is known to be $O(\tau)$.

We highlight the benefits of the two-run extrapolation procedure with
an example calculation on the 1D HEG\@. Once the maximum allowed time
step $\tau_{\mathrm{max}}$ in the linear regime had been determined,
pairs of runs were performed at $\tau_{2}=\tau_{\mathrm{max}}$ and
incrementally smaller time steps $\tau_{1}$. The pairs of runs were
each performed using the same total amount of computing time. The time
was distributed either to ensure equal-sized error bars or according
to the prescription $T_1/T_2= (\tau_2/\tau_1)^{3/2}$ to guarantee
minimal final extrapolated error. The simulation times were sufficient
to ensure that the data could be reblocked for accurate error
estimates. The final extrapolated energy estimates all agreed to
within the expected uncertainty, consistent with the assertion that
all of the time steps are within the linear regime. The results shown
in \figref{fig:TimestepExtrapolation} highlight that, for the range of
$\tau_2/\tau_1$ tested, there is strong agreement between the
analytical prediction and the DMC results. In particular, the error
bar on the extrapolate with the optimal distribution of effort is
clearly minimized by the choice $\tau_2/\tau_1=4$. The distribution of
effort according to $T_1/T_2= (\tau_2/\tau_1)^{3/2}$ yields a modest
computational advantage over the choice $T_1/T_2=\tau_2/\tau_1$.

In summary, to minimize the statistical error bar on the DMC energy
extrapolated to zero time step, one should perform one DMC calculation
at the largest time step $\tau_{\rm max}$ for which the bias is still
linear in the time step and a second DMC calculation with time step
$\tau_{\rm max}/4$.  Eight times as much computational effort should
be devoted to the latter calculation as to the former.  One could use
a similar approach to optimize the efficiency of extrapolating to
infinite population or to infinite system size in a QMC study of
condensed-matter systems.


\section{Reblocking\label{sec:reblocking}}
The use of small time steps in DMC results in serially-correlated
data. For accurate estimates of the statistical uncertainties of DMC
expectation values, the serial correlation must be accounted
for. Here, we investigate reblocking~\cite{flyvbjerg-eeoaocd1989},
which is advantageous due to its computational convenience and ease of
implementation. We propose a scheme for the choice of block length
such that accurate error bars may be reliably determined when an
estimate for the correlation length is unavailable and must be
obtained directly from the data.


For most random processes used in Monte Carlo methods the serial
correlation is purely positive, so that the standard error (treating
all samples as independent) should be multiplied by an error factor
$\eta_{\mathrm{{err}}} \geq 1$. Let the new estimate of the standard
error be $\Delta$, and let $\nind$ be $n$ divided by the
\textit{estimated} correlation length, \textit{i.e.,} $\nind\leq n$
measures the estimated effective number of steps. We may express
$\Delta$ as
\begin{equation}
  \Delta = \eta_{\mathrm{{err}}} \sqrt{{\mathrm{var}}[E_i] / n} = 
\sqrt{{\mathrm{var}}[E_i] /
    \nind},\label{eq:errfac}
\end{equation}
where $\mathrm{var}[E_i]$ is the sample variance of the $n$ data
points $\lbrace E_i \rbrace$ and the error factor
$\eta_{\mathrm{{err}}}$ is the square root of the estimated
correlation length~\cite{wolff-mcewle2004}.  As each step of a QMC
calculation is associated with a time step $\tau$ measured in physical
units, a correlation time in physical units can be defined as $\tcorr
= \tau \ncorr(\tau)$.  In the limit $\tau \rightarrow 0$, the
integrated correlation time $\tcorr$ becomes independent of $\tau$ and
takes a value characteristic of the system under study.

To estimate $\eta_{\mathrm{{err}}}$ from a set of data points $E_i$, there are
several commonly-used approaches: computing the correlation length,
reblocking, or using resampling techniques like the jackknife and bootstrap
methods~{\cite{wolff-mcewle2004,flyvbjerg-eeoaocd1989,jun1995jackknife,chernick2008bootstrap}}. Here
we have focused on the reblocking method because it is computationally
convenient (and conceptually very simple) to apply reblocking continuously as
local observable data are appended to the stored results, vastly reducing
memory requirements~{\cite{kent-eafoeaolodscd2007}}. A naive calculation of
the correlation-corrected statistical error necessitates the storage of $O(n)$
observable values, whereas reblocking on-the-fly reduces this to $O[\log(n)]$.

Reblocking is a method in which a sequence of $n$ serially correlated
data points is divided into contiguous blocks of length $B$, and the
raw data are averaged within each of these blocks, defining a new data
set of length $n/B$. The naive variance of the reblocked estimate of
the mean is larger than that of the original data, although the mean
itself is unchanged. The estimated error initially increases with $B$,
reaching a plateau once the serial correlation has approximately been
removed from the data. When $B$ approaches $n$, the plot becomes very
noisy due to the small number of blocks.

\begin{figure}
 \centerline{\resizebox{8.5cm}{!}{\includegraphics{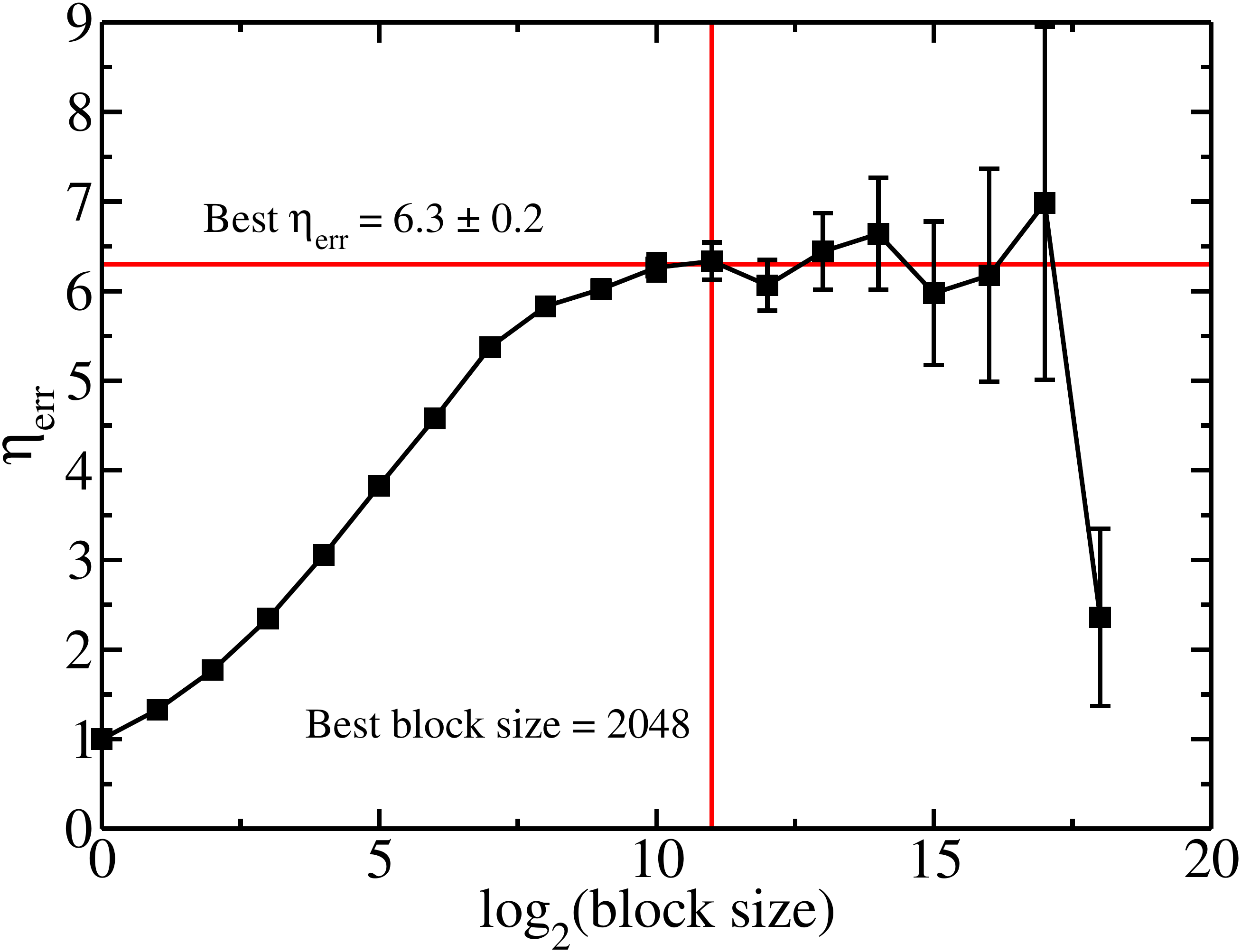}}}
 \caption{(Color online) Reblocking analysis of a typical DMC run (Li atom,
with $\tau = 0.01$ a.u.\ and 900000 time steps). The optimal block size is
chosen by the algorithm described in the text.}
 \label{fig:reblock}
\end{figure}

The reblocking analysis of a typical DMC run is shown in Fig.\
\ref{fig:reblock}.  The fundamental difficulty in interpreting this
kind of data is the choice of an appropriate block size. In the case
presented here, the run time of 900000 time steps was sufficiently
long to form a clear plateau in the reblock plot. However,
individually inspecting the reblocked data of each calculation to make
a choice by eye is neither objective nor efficient. Table
\ref{table:reblock_ncorr} shows the estimated correlation lengths from
reblocking the Li data with different block lengths.

\begin{table}[!ht]
\begin{center}
\begin{tabular}{l@{\hspace{1cm}}l}
  \hline \hline
  $\log_2 B$
& Estimated $n_{\mathrm{corr}}$
 \\
  \hline
1  &  1.762(4) \\
2  &  3.140(9) \\
3  &  5.51(2)  \\
4  &  9.34(5)  \\
5  &  14.7(1)  \\
6  &  21.0(2)  \\
7  &  28.9(5)  \\
8  &  34.0(8)  \\
9  &  36(1)    \\
10 &  39(2)    \\
11 &  40(3)    \\
12 &  37(3)    \\
13 &  42(6)    \\
14 &  44(8)    \\
  \hline \hline
\end{tabular}
\caption{ The estimated correlation length found from reblocking DMC data
 with block size $B$. The system was the Li atom with $\tau = 0.01$ a.u.\
  and 900000 time steps.
  \label{table:reblock_ncorr}}
\end{center}
\end{table}

A simple yet robust algorithm for automatically choosing the best
block size is as follows. Following Ref.\ \onlinecite{wolff-mcewle2004}, the
block size
\begin{equation}
  B_{\mathrm{{opt}}} = \sqrt[3]{2
  n \ncorr^2}
\end{equation}
offers an appropriate balance between the systematic and the
statistical error in the estimate of the standard error for any set of
$n$ data points with the integrated correlation length $\ncorr$. If a
good estimate for $\ncorr$ is available before the data are analyzed,
it is best to use this and thereby make the choice of the block size
independent of the statistical data themselves. In many studies,
several runs on similar systems are needed or the knowledge of
$\tcorr$ can be used to extrapolate $\ncorr$ to small time steps. In
such cases it is best to estimate $\ncorr$ once and reuse it for the
choice of $B_{\mathrm{{opt}}}$ in subsequent calculations, provided
that the physical system and wave-function quality (and thus the
correlation length) are unchanged. The error factor
$\eta_{\mathrm{{err}}}$ obtained in each case can then be used to
double-check the transferability of the estimated correlation length
without influencing the choice of $B_{\mathrm{{opt}}}$, so that there
is no bias from manually making a data-dependent choice.

If an independent estimate for $\ncorr$ is not available, it has to be
obtained from the analyzed data themselves. In this case, the
estimated correlation length $\eta_{\mathrm{{err}}}^2$ depends on the
choice of $B$, so the condition for $B_{\mathrm{{opt}}}$ becomes
recursive. We consider block sizes that are powers of $2$ and start
with the largest block size possible, decreasing $B$ and examining the
error factor. The optimal block size is then the last value of $B$ for
which the inequality $B^3 > 2 n \eta_{\mathrm{{err}}}^4 (B)$ is
satisfied. We may restrict $B$ to powers of $2$ since the block length
is expected to be logarithmically distributed.

In a reblocking analysis for $n$ data points, the relative error in the
error factor for a given block size depends only on the number of
blocks as
\begin{equation}
  \frac{\delta \eta_{\mathrm{{err}}} \left( B
  \right)}{\eta_{\mathrm{{err}}} \left( B \right)} =
  \sqrt{\frac{B}{2 n}} .
\end{equation}
Assuming that a user would typically expect at least one significant digit in
the standard error, we can further define a straightforward criterion for the
success of a reblocking analysis: if $B_{\mathrm{{opt}}} < n / 50$, the
analysis can be accepted as successful, otherwise the reliability of the
result is questionable and one should gather more data. Except for systems
with distinct correlation times at extremely different scales, this criterion
is expected to be reliable in all typical cases occurring in QMC\@.  More than
one correlation time might occur in weakly bound molecules; the longest
correlation time is defined by the size of the molecule and the shortest is
determined by the Bohr radius of the nucleus with the highest atomic
number. In such cases, it may be necessary to accumulate more data; the block
size should clearly be determined by the longest correlation length.

In summary, when using reblocking to remove serial correlation from
QMC data, one should ideally obtain an accurate estimate of the
correlation length separate from the data being analyzed and use
$B_{\mathrm{{opt}}} = \sqrt[3]{2 n \ncorr^2}$ to determine the block
length~\cite{wolff-mcewle2004}. If this is not possible, one should
aim to satisfy the inequalities $B^3 > 2 n \eta_{\mathrm{{err}}}^4
(B)$ and $B_{\mathrm{{opt}}} < n / 50$ for a reliable and accurate
estimate of the error.

All methods of accurately calculating the error bar from
serially-correlated data implicitly estimate the correlation
length. The noise and associated uncertainty in estimates of the
correlation length introduce error into the estimated statistical
error bar. In the next section we describe how this can increase the
apparent number of outlying results.

\section{Outliers in QMC results\label{sec:outliers}}
\subsection{Introduction}

In this section, we investigate the frequency with which ``outliers''
occur in QMC results.  We define an outlier as a result located more
than a given number of \textit{estimated} error bars from the
underlying mean value. For example, one may fit a straight line to DMC
energies at small $\tau$. If there are sufficient data points, the
linear fit is a good estimate of the underlying mean; one would
usually expect, by the central limit theorem (CLT), a fraction $0.32$
of the points to deviate from the fitted function by more than a
single error bar. Here we address the observation that QMC estimates
can lie outside statistical error bars of the underlying mean more
often than one would expect were the error bars correctly describing
the width of an underlying Gaussian distribution. We will demonstrate
that uncertainty in the estimated correlation length is largely
responsible for this effect.

We begin with direct observation of the numbers of outliers for two
systems, the C atom and the Si crystal. By performing a large number
of short VMC calculations for each system, we count directly the
number of energies occurring more than $Q$ error bars from the
underlying mean, where the error is estimated separately for each
run. Each estimate of the statistical error is also implicitly an
estimate of the correlation length, as described by Eq.\
(\ref{eq:errfac}).

To complement the direct approach, we then derive an analytic
expression for the fraction of points expected to lie more than $Q$
error bars from the mean under the assumption that the distribution of
local energies is Gaussian. The resulting expression depends on the
distribution of estimated correlation lengths. Finally, we compare the
expected result from this purely Gaussian model process with that
found earlier from VMC, forming conclusions about the validity of the
Gaussian assumption and the origin of outliers.

\subsection{VMC calculations\label{sec:vmc_outliers}}

We have performed a large number of VMC calculations for two typical
systems; the all-electron carbon atom and a periodic crystalline
silicon system. For the C atom we performed $5\times 10^4$, $2\times
10^4$, and $10^4$ calculations of length $200$, $500$, and $1000$
steps, respectively. The Si system used a periodic simulation cell
containing 54 silicon atoms, where the $1s^2 2s^2 2p^6$ electrons are
described by pseudopotentials. For the Si system, we performed
$1.5\times 10^5$, $7.5\times 10^4$, and $3\times 10^4$ calculations of
length $100$, $200$, and $500$ steps, respectively.

A short calculation yields an energy and estimated error. From the
data we estimate the probability ${\rm P}\left
  (\delta\bar{E}>Q\Delta\right )$ of observing a VMC energy $\bar{E}$
at a position more than $Q\Delta$ from the true mean $E_0$, where
$\delta\bar{E}=|\bar{E}-E_0|$ and $\Delta$ is the \textit{estimated}
error bar, itself also a random variable. The underlying mean $E_0$ is
calculated accurately using a much longer run. If the error bars
exactly described the width of an underlying Gaussian distribution,
one would expect ${\rm P}\left (\delta\bar{E}>Q\Delta\right
)=\mathrm{erfc}(Q/\sqrt{2})$.  The symbols in Figs.\
\ref{fig:silicon_erfc_modif} and \ref{fig:carbon_erfc_modif} show the
deviation of the VMC results from this ideal case.

\begin{figure}
\includegraphics[scale=0.3]{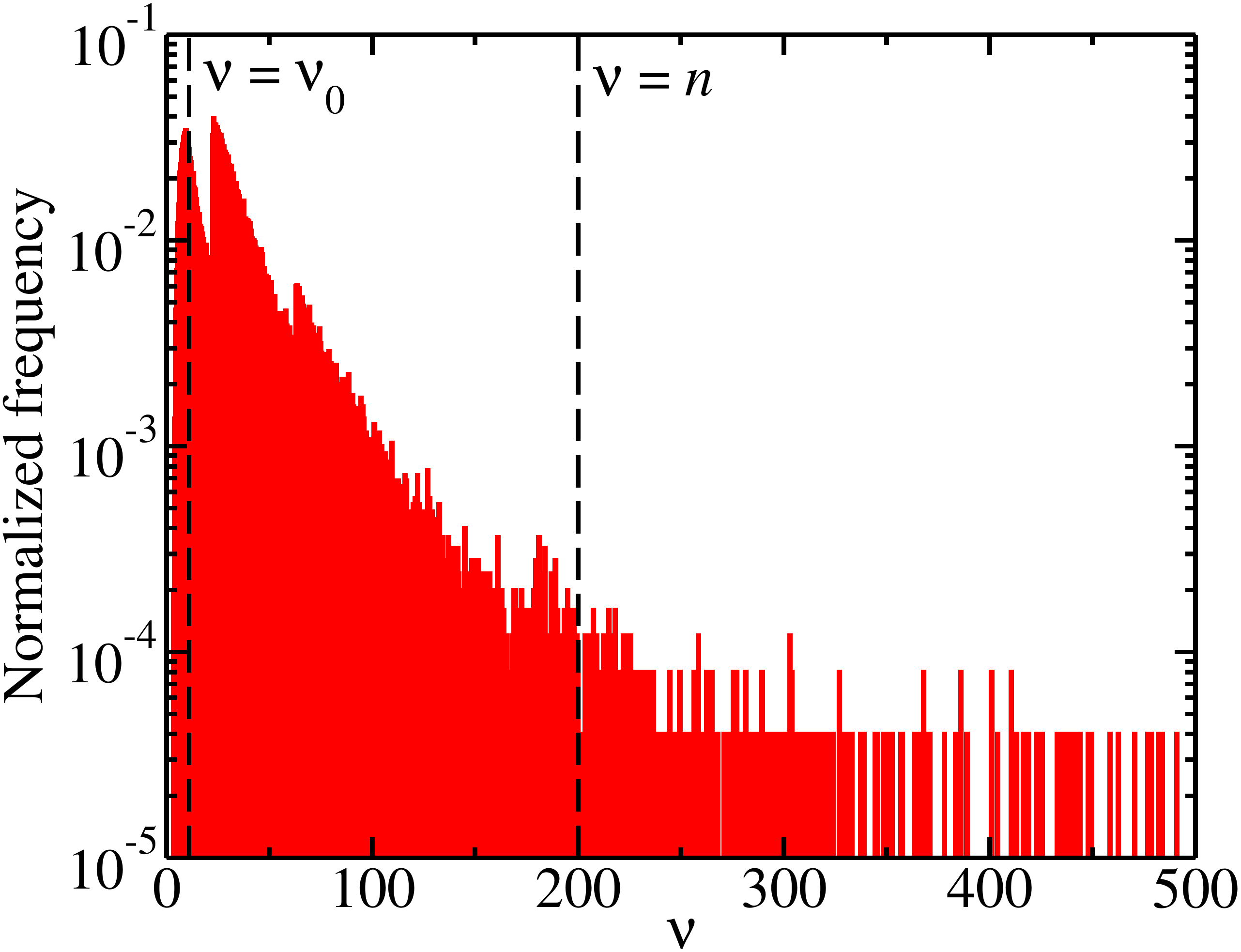}
\caption{(Color online) Distribution of
  $\nu=n/\eta^2_{\mathrm{{err}}}$ from performing $5\times 10^4$
  all-electron VMC calculations for the C atom. Each calculation
  consisted of $n=200$ steps and the error factors were obtained by
  reblocking. The dashed lines show the accurate effective number of
  steps, $\nind_0$, and the effective number of steps corresponding to
  no serial correlation,
  $\nind=n$.\label{fig:carbon_dist_n_over_ncorr}}
\end{figure}

By estimating the statistical error bar for each run, we are able to
estimate ${\rm p}_{\mathrm{ind}}$, which is the distribution of the
estimated effective number of steps $\nu=n/\eta^2_{\mathrm{{err}}}$,
where $n$ is the number of VMC steps and $\eta_{\mathrm{err}}$ is the
error factor of Eq.\ (\ref{eq:errfac}). An example kernel estimate of
${\rm p}_{\mathrm{ind}}$ is shown in Fig.\
\ref{fig:carbon_dist_n_over_ncorr}; one can see that $\nind$ is
occasionally larger than $n$. This is clearly unphysical, stemming
from noise in the estimate of the correlation length, and results in
underestimation of the statistical error bar. The distribution ${\rm
  p}_{\mathrm{ind}}$ appears to decay at large $\nu$ as $\nu^{-A}$,
where $A$ is between $4.5$ and $6.5$.

\subsection{Gaussian model}

We now attempt to replace VMC sampling with an ideal process where the
underlying distributions are Gaussian.  Our starting point is the
distribution of local energies, ${\rm p}_{\mathrm{loc}}$, from which
energies are drawn at successive points along the random walk in
configuration space. The quantity of interest is again the probability
${\rm P}\left (\delta\bar{E}>Q\Delta\right )$ of observing a sample
mean energy $\bar{E}$ at a position more than $Q\Delta$ from the true
mean $E_0$.

Let us assume that the distribution of local energies is Gaussian,
\begin{equation}
   {\rm p}_{\mathrm{loc}}(\Eloc)=\frac{1}{\sqrt{2\pi}\sigma_{0}}\exp \left (
    \frac{-(\Eloc-E_0)^2}{2\sigma_{0}^2} \right )\;,
\label{eq:gaussian_loc_e}
\end{equation}
where $\sigma_{0}^2$ is the variance of the distribution. Consider
drawing $n$ samples $\lbrace E_i\rbrace_{i=1,\ldots,n}$ from the PDF
of Eq.\ (\ref{eq:gaussian_loc_e}) using the Metropolis algorithm; this
yields $\nu_0\leq n$ independent samples due to serial correlation.
For this simple case the sample mean, $\bar{E}=(1/n) \sum^n_{i=1}
E_i$, has the distribution
\begin{equation}
    {\rm p}_{\mathrm{ave}}(\bar{E})=\sqrt{\frac{\nu_0}{2\pi\sigma_{0}^2}}
    \exp \left (\frac{-(\bar{E}-E_0)^2}{2\sigma_{0}^2/\nu_0} \right )\;.
\label{eq:gaussian_mean_e}
\end{equation}
The statistical error bar on $\bar{E}$ is calculated from the same set
of local energies as the estimate itself.  However, since estimates of
the correlation length are subject to noise, there is uncertainty in
the effective number of independent samples. Although this leaves
$\bar{E}$ unaffected, it does influence the estimated error. As
before, we define $\nind$ as the random estimate of $\nu_0$ and again
refer to the PDF ${\rm p}_{\mathrm{ind}}$ from which $\nind$ is drawn.

It is well-known that a sum of squares of normally-distributed random
numbers follows the chi-square
distribution~\cite{cochran1934distribution}. Since the error bar
$\Delta$ is related to the sample variance through Eq.\
(\ref{eq:errfac}), we can write down the bivariate PDF ${\rm
  p}_{\mathrm{err}}$ for $\Delta$ and $\nind$,
\begin{equation}
{\rm p}_{\mathrm{err}}(\Delta,\nind)=
\frac{\Delta^{\nind-2}
\exp \left [ -\frac{\nind(\nind-1)\Delta^2}{2\sigma_0^2}\right ]{\rm p}_{\mathrm{ind}}(\nind)}
{\left ( \frac{\nind(\nind-1)}{\sigma_0^2}\right )^{\frac{1-\nind}{2}}
2^{\frac{\nind-3}{2}}\;\Gamma\left (
\frac{\nind-1}{2}\right )}\;,
\label{eq:gaussian_err_dist}
\end{equation}
where $\Delta$ is only allowed to take positive values and $\Gamma$ is
the Gamma function.
It is straightforward to find analytically the probability of
observing an energy more than $Q$ error bars from the mean as a
function of $Q$ and $\Delta$. This is done by integrating Eq.\
(\ref{eq:gaussian_mean_e}),
\begin{equation}
2\int_{E_0+Q\Delta}^{\infty} {\rm d}\bar{E}\;{\rm p}_{\mathrm{ave}}(\bar{E})
={\rm erfc}\left (
  \frac{Q\Delta}{\sigma_0}\sqrt{\frac{\nu_0}{2}}
\right )\;.
  \label{eq:energy_integral}  
\end{equation}
To find the desired probability, ${\rm P}\left
  (\delta\bar{E}>Q\Delta\right )$, we evaluate the expectation value
of Eq.\ (\ref{eq:energy_integral}) with respect to the distribution of
$\Delta$ and $\nind$,
\begin{eqnarray}
{\rm P}\left (\delta\bar{E}>Q\Delta\right ) & = &
\int_2^\infty {\rm d}\nind
\int_0^\infty {\rm d}\Delta\;
{\rm p}_{\mathrm{err}}(\Delta,\nind)
\nonumber \\
& \times&
{\rm erfc}\left (
  \frac{Q\Delta}{\sigma_0}\sqrt{\frac{\nu_0}{2}}
\right )
\;,
\label{eq:prob_integral}
\end{eqnarray}
where we have used the fact that the sample mean and sample variance
are independent for Gaussian distributed random
variables~\cite{studentsratio,basustheorem}. To evaluate the integral
of Eq.\ (\ref{eq:prob_integral}), we require the distribution ${\rm
  p}_{\mathrm{ind}}$ and an accurate estimate of the true effective
number of steps, $\nu_0$. We will take these quantities from the VMC
results of Sec.\ \ref{sec:vmc_outliers}, so that the integral of Eq.\
(\ref{eq:prob_integral}) represents an ideal Gaussian process
accompanied by the uncertainty in the number of independent samples
(and thus the correlation length) that we observe in VMC\@. The
integral of Eq.\ (\ref{eq:prob_integral}) can then be evaluated
numerically.

\subsection{Results}

\begin{figure}
  \includegraphics[scale=0.32]{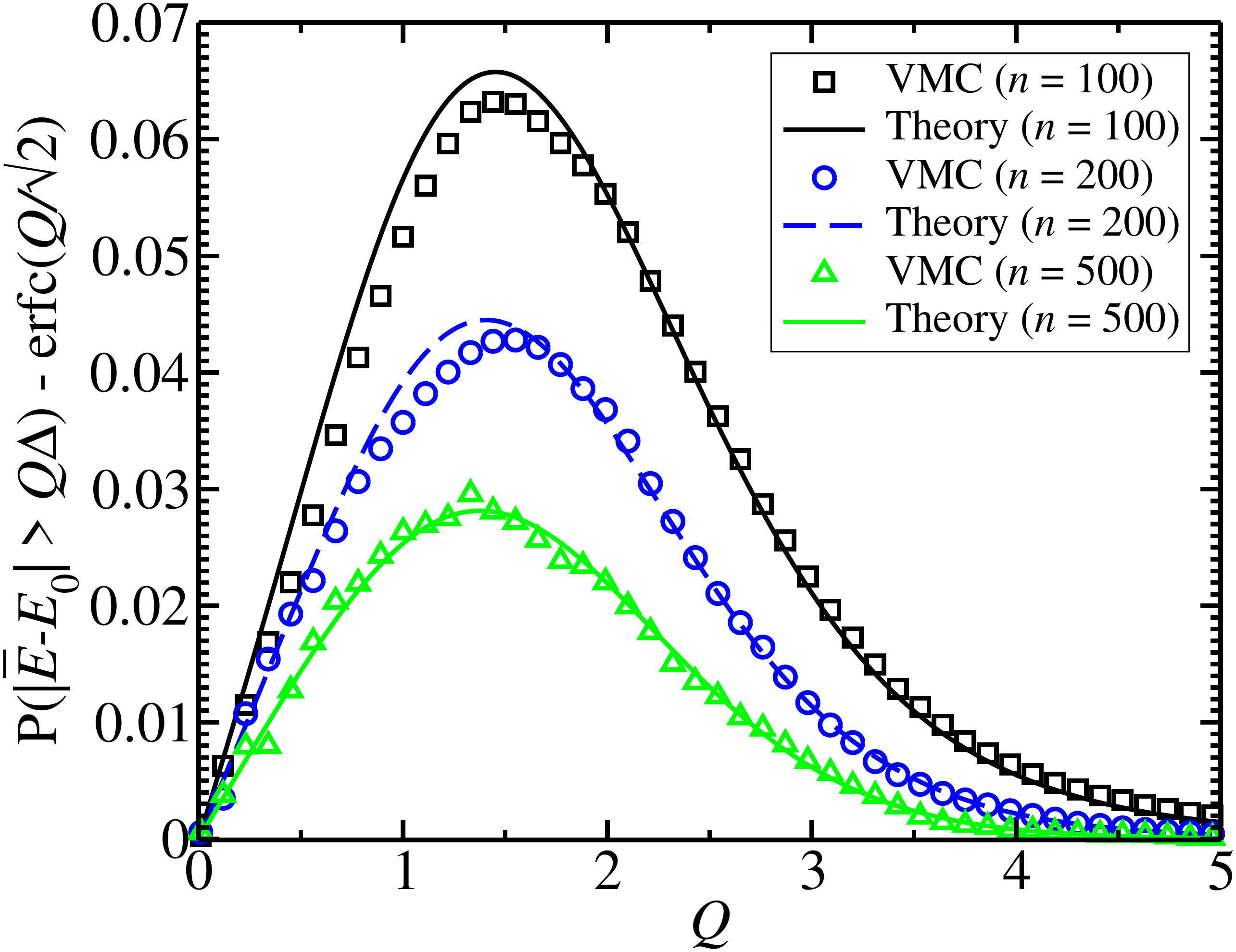}
  \caption{(Color online) Enhancement to the probability of observing an
    energy more than $Q$ error bars from the mean for 54-atom
    (216-electron) bulk Si. The square, circular and triangular symbols 
    show the results of VMC calculations of $n=100$, $200$ and $500$ local 
    energies, 
    respectively. 
    The number of calculations for each set was $(1.5\times 10^7)/n$.
    The lines show the results of evaluating the
    integral of Eq.\ (\ref{eq:prob_integral}), where $\nu_0$ and ${\rm
      p}_{\mathrm{ind}}$ were determined from the VMC data.
    \label{fig:silicon_erfc_modif}}
\end{figure}
\begin{figure}
  \includegraphics[scale=0.32]{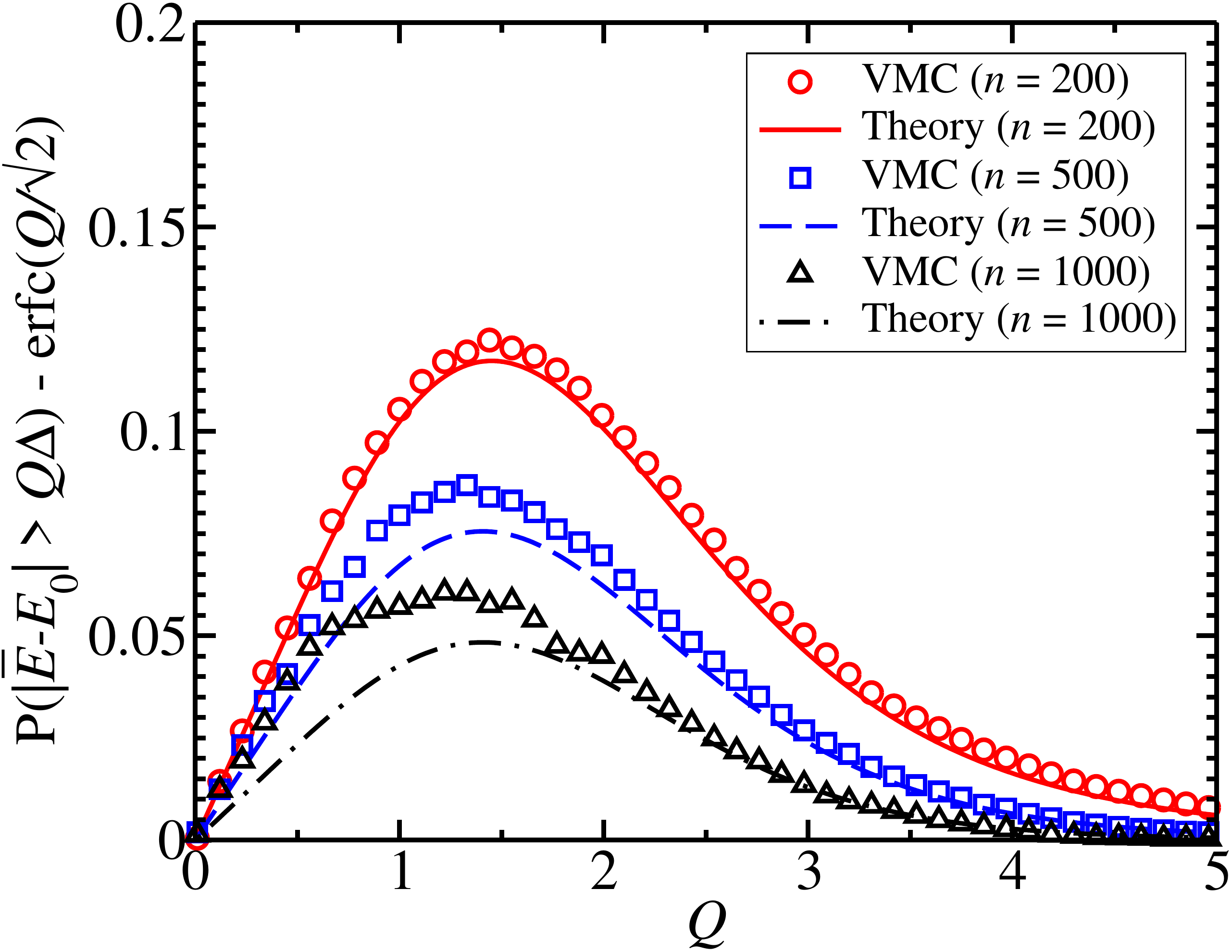}
\caption{(Color online) Enhancement to the probability of observing an energy
  more than $Q$ error bars from the mean for the C atom. The circles, squares
  and triangles represent all-electron VMC calculations with $n=200$, $500$,
  and $1000$ local energies, respectively.  The number of calculations for
  each set was $10^7/n$. The lines represent the results of evaluating the
  integral of Eq.\ (\ref{eq:prob_integral}), where $\nu_0$ and ${\rm
  p}_{\mathrm{ind}}$ were determined from the VMC data.
    \label{fig:carbon_erfc_modif}}
\end{figure}

Figures \ref{fig:silicon_erfc_modif} and \ref{fig:carbon_erfc_modif}
show the actual fractions of outliers from the VMC calculations
compared with those predicted by Eq.\ (\ref{eq:prob_integral}), which
used ${\rm p}_{\mathrm{ind}}$ and $\nu_0$ from the VMC calculations
but otherwise assumed a model Gaussian process. The fraction of points
occurring more than $Q$ error bars from the mean has been offset by
$\mathrm{erfc}(Q/\sqrt{2})$ in the figures to highlight the deviation
from the result when the correlation length is known exactly,
\textit{i.e.,} {${\rm p}_{\mathrm{ind}}(\nind)=\delta(\nind-\nu_0)$}.

When $n$ takes smaller values, the uncertainty in the correlation
length is greater and the fraction of points which may be classified
as outliers is larger. A poor trial wave function could also
contribute to the effect by reducing the sampling efficiency. In the
case of the C atom, instead of the $0.13$ probability of observing an
energy more than $1.5$ error bars from the mean that one would expect
on the basis of Gaussian statistics, the VMC results are consistent
with a $0.25$ probability (for runs of $200$ local energies). For the
C and Si systems, estimating the error bars for each short run using a
single more accurate estimate of the correlation length (from a single
longer run or by averaging the estimates from each shorter run),
results in a return to {${\rm P}\left (\delta\bar{E}>Q\Delta\right
  )=\mathrm{erfc}(Q/\sqrt{2})$}.

For systems exhibiting singularities in the local energy, the CLT
converges only very slowly and one might expect the non-Gaussian
character of $\rm p_{\Eloc}$ to play a role in determining the
frequency with which outliers are
observed~\cite{trail-hreiqmc2008}. Singularities in the local energy
arise when the description of the wave-function nodes is inexact, as
is the case for the C and Si systems considered here, and when the
cusp conditions are unfulfilled.


We find that the contribution from the non-Gaussian parts of the
energy PDF towards the frequency of outliers is statistically
insignificant. The evidence for this is twofold; first, the integrals
based on a purely Gaussian ${\rm p}_{\mathrm{loc}}$ agree very well
with the VMC data, suggesting that uncertainty in the correlation
length is almost solely responsible for the effect. Secondly,
attempting to fit a function with power law tails (of the form
suggested in Ref.\ \onlinecite{trail-hreiqmc2008}) to the VMC energies
yields very small values for the weight under the tails (usually
within error bars of zero), even though the distribution of local
energies is itself manifestly non-Gaussian~\footnote{We form a biased
  estimate for the weight of the power-law tails by fitting Eq.\ (48)
  of Ref.\ \onlinecite{trail-hreiqmc2008} to the distribution of
  energies obtained from $10^4$ VMC runs, each of $1000$ steps. We
  find $\lambda_3=1.1(8)$ and $\lambda_3=0.2(4)$ for the C atom and
  the bulk Si system, respectively. The $\chi^2$ error in the fit was
  $0.95$ per data point for the C atom and $1.03$ per data point for
  the bulk Si system}.

In conclusion, when there is too little data to make an accurate
estimate of the correlation length, the estimated error is subject to
an uncertainty that increases the probability of observing
outliers. For isolated calculations of a single run, the problem
amounts to the gathering of sufficient data for an accurate estimate
of the correlation length. Where dependence upon several parameters is
being investigated for large systems, one should calculate accurately
the correlation length from a single long run or by averaging many
estimates from shorter runs. The accurate estimate of the correlation
length can then be interpreted as the square of the error factor,
$\eta_{\mathrm{err}}$, and used to calculate the error bars on related
calculations in two ways: either by guiding the choice of block length
($B^3=2n\eta_{\mathrm{err}}^4$) or by multiplying the unreblocked
error by $\eta_{\mathrm{err}}$; the two estimates should be roughly
consistent.

\section{Conclusions\label{sec:conclusions}}

In this paper we have developed and carefully tested new ways of
improving the efficiency of QMC calculations.

Our analysis of VMC efficiency shows that the use of decorrelation
loops approximately doubles the efficiency of EBES, with a loop of
three moves providing the greatest benefit for a wide range of
systems. The improvement in efficiency for CBCS is much
greater. However, we find that EBES rather than CBCS yields a higher
efficiency, except in small systems where backflow transformations are
used. Of the automatic schemes for optimizing the time step that we
have considered, attempting to achieve a move acceptance ratio of
$50\%$ leads to the greatest efficiency within EBES\@.

For the extrapolation of DMC energies to zero time step there is a
clear optimal strategy. One must first find the largest time step
$\tau_2$ for which the energy can be considered to vary linearly with
time step. One should then minimize the error in the extrapolate by
performing calculations at two different time steps; the first at
$\tau_1=\tau_2/4$ with computational effort $8T/9$, and the second at
$\tau_2$ with computational effort $T/9$, where $T$ is the total
computing time available.

The reblocking method of removing serial correlation from QMC data
offers a significant computational advantage over other
methods. Ideally, when choosing a block size, one should estimate the
correlation length for a system independently of the
serially-correlated data themselves. The optimal block length $B$
should be chosen such that $B^3>2n\eta_{\mathrm{err}}^4$ and $B<n/50$
[where $n$ is the number of data points and $\eta_{\mathrm{err}}$ is
the error factor of Eq.\ (\ref{eq:errfac})]. This allows automated
data processing with a warning criterion for insufficient data that
works reliably in the absence of multiple correlation periods
occurring on distinctly different scales.

Finally, we note that uncertainty in the correlation length leads to
estimated error bars that have the potential to increase the
probability of observing outliers in QMC results. The size of the
effect is dependent on the system and wave function. One can alleviate
the problem by calculating the statistical error using an accurate
estimate of the correlation length from a longer run. Otherwise, our
findings highlight the importance of sufficient statistics-gathering
and caution when interpreting DMC results for large systems.

Quantum Monte Carlo techniques are not as widely used as other methods
due to their computational expense and the complexity of carrying out
a calculation. In addition to improving the statistical and
computational efficiency of QMC calculations, the strategies we have
described are straightforward to automate. With the implementation of
such schemes, QMC has the potential to evolve into a true black-box
tool. This will facilitate wider use of the method and improve its
reliability.

\section{Acknowledgments}
The authors thank John Trail and Richard Needs for many helpful
conversations. RML is grateful for the support of the Engineering and
Physical Sciences research council (EPSRC) of the UK\@. GJC acknowledges
the support of the Royal Commission for the Exhibition of 1851, the
Kreitman Foundation, the Feinberg Graduate School and the National
Science Foundation under Grant No.\ NSF PHY05-51164.  NN thanks the
DAAD, the EPSRC and the HECToR dCSE programme, and NDD acknowledges
support from the Leverhulme Trust.

\end{document}